\documentclass[structabstract]{aa}  
\newcommand{\Ms}{$M_{\odot}$}

\newcommand{\rstar}{R$_{\star}$}
\newcommand{\ace}{C$_2$H$_2$}
\newcommand{\prop}{C$_3$H$_3$}
\newcommand{\ben}{C$_6$H$_6$}

\newcommand{\coro}{C$_{24}$H$_{12}$}
\newcommand{\micron}{$\mu$m}
\usepackage{graphicx}
\usepackage{color}
\usepackage{txfonts}
\begin{document}
   \title{The inner wind of IRC+10216 revisited: New exotic chemistry and diagnostic for dust condensation in carbon stars}

%   \subtitle{I. Overviewing the $\kappa$-mechanism}

   \author{I. Cherchneff
%          \inst{1}
%       \and
%         C. Ptolemy\inst{2}\fnmsep\thanks{Just to show the usage
%      of the elements in the author field}
          }

   \institute{Departement Physik, Universit\"{a}t Basel, Klingelbergstrasse 82, 4056 Basel, Switzerland\\
              \email{isabelle.cherchneff@unibas.ch}
%         \and
%             University of Alexandria, Department of Geography, ...\\
%             \email{c.ptolemy@hipparch.uheaven.space}
%             \thanks{The university of heaven temporarily does not
%                     accept e-mails}
             }

   \date{Received November 29, 2011; accepted April 26, 2012}

% \abstract{}{}{}{}{} 
% 5 {} token are mandatory
 
  \abstract
  %context heading (optional)
 {}
   {We model the chemistry of the inner wind of the carbon star IRC+10216 and consider the effects of periodic shocks induced by the stellar pulsation on the gas to follow the non-equilibrium chemistry in the shocked gas layers. We consider a very complete set of chemical families, including hydrocarbons and aromatics, hydrides, halogens, and phosphorous-bearing species. Our derived abundances are compared to those for the latest observational data from large surveys and the Herschel telescope.}
  {A semi-analytical formalism based on parameterised fluid equations is used to describe the gas density, velocity, and temperature from 1 \rstar~to 5 \rstar. The chemistry is described using a  chemical kinetic network of reactions and a set of stiff, ordinary, coupled differential equations is solved.   }
   {The shocks induce an active non-equilibrium chemistry in the dust formation zone of IRC+10216 where the collision destruction of CO in the post-shock gas triggers the formation of O-bearing species such as  H$_2$O and SiO. Most of the modelled molecular abundances agree very  well with the latest values derived from Herschel data on IRC+10216. The hydrides form a family of abundant species that are expelled into the intermediate envelope. In particular, HF traps all the atomic fluorine in the dust formation zone. The halogens are also abundant and their chemistry is independent of the C/O ratio of the star. Therefore, HCl and other Cl-bearing species should also be present in the inner wind of O-rich AGB or supergiant stars. We identify a specific region ranging from 2.5 \rstar~to 4 \rstar, where polycyclic aromatic hydrocarbons form and grow. The estimated carbon dust-to-gas mass ratio derived from the mass of aromatics formed ranges from $1.2 \times 10^{-3}$ to $5.8 \times 10^{-3}$ and agrees well with existing values deduced from observations. This aromatic formation region is situated outside hot layers where SiC$_2$ is produced as a bi-product of silicon carbide dust synthesis. The MgS grains can form from the gas phase but in lower quantities than those necessary to reproduce the strength of the 30 \micron~emission band. Finally, we predict that some molecular lines will show a flux variation with pulsation phase and time (e.g., H$_2$O), while other species will not (e.g., CO). These variations merely reflect the non-equilibrium chemistry that destroys and reforms molecules over a pulsation period in the shocked gas of the dust formation zone.}
   {}

   \keywords{Stars: Carbon --
                Astrochemistry --
                Stars: AGB and post-AGB --
                Molecular processes
               }
\authorrunning{I. Cherchneff}
\titlerunning{The dust formation zone of IRC+10216 revisited}
   \maketitle
%
%________________________________________________________________

\section{Introduction}

In their late stages of evolution, low-mass stars (i.e., stars with initial masses on the Zero-Age-Main-Sequence comprised between 1 and 8 \Ms) ascend the Asymptotic Giant Branch (AGB) and develop cool and strong stellar winds characterised by a great variety of chemical species in the outflow detected through their ro-vibrational transitions (Olofsson 2008). With the launch of the submillimetre (submm) Herschel telescope and the beginning of science operations of the Atacama Large Millimetre Array (ALMA), our knowledge of the chemical composition of AGB winds is bound to dramatically increase with the discovery and identification of many new molecules. The wind of a AGB star develops in the dense and hot gas layers above the stellar photosphere, triggered by the formation of dust. To a first approximation, the chemical composition of the photosphere is determined by thermodynamic equilibrium (TE) owing to the high temperatures and densities of the gas (Tsuji 1973, McCabe et al. 1979). Carbon monoxide, CO, which is a very stable species, forms under TE conditions after the production of molecular hydrogen, H$_2$. If the stellar photosphere is oxygen-rich, the excess oxygen not locked up in CO drives a wind chemistry dominated by oxygen-bearing molecules such as water, H$_2$O, and silicon monoxide, SiO, and the dust forming in the wind includes silicates (e.g., forsterite Mg$_2$SiO$_4$) and metal oxides (e.g., alumina Al$_2$O$_3$). Conversely, if the star has experienced third-dredge up in the upper part of its AGB ascension, its photosphere may become carbon-rich. The chemistry of the wind then reflects the excess carbon not locked up in CO and is rich in C-bearing species such as acetylene, C$_2$H$_2$, and hydrogen cyanide, HCN. Those stars form large amounts of carbon dust close to their photosphere. This simple picture was seriously questioned by the detection at millimetre (mm) wavelengths of HCN in O-rich AGB stars (Deguchi \& Goldsmith 1985). In carbon stars, SiO was observed (Olofsson et al. 1982), but because TE models of carbon stars predicted its formation, it did not come as a surprise to observe this O-bearing species in carbon-rich environments. However, the observed abundances were much higher than those derived from TE. Several mechanisms were proposed to explain these unexpected species, including an ion-molecule chemistry in the outer part of the stellar wind experiencing the penetration of the ultraviolet (UV) interstellar radiation field and cosmic rays (Nejad \& Millar 1982). The later detection by the Infrared Space Observatory, ISO, of hot vibrational transitions of carbon dioxide, CO$_2$, in the cool supergiant NML Cyg (Justtanont et al. 1996) indicated that carbon-bearing species could form in the deep layers of O-rich stellar winds. The detection of water, H$_2$O, with the SWAS satellite at submm wavelength by Melnick et al. (2001) and that of hydroxyl, OH, by Ford et al. (2004) in the carbon star IRC+10216 provided additional evidence of the complex chemistry of AGB outflows, where comets (Melnick et al. 2001), grain-surface chemistry (Willacy 2004), or UV photodissociation in the intermediate envelope (Ag{\'u}ndez \& Cernicharo 2006) were proposed as possible sources of water. 

However, that these unexpected species could form by means of some non-equilibrium chemistry in the inner wind was soon found to be viable by surveys of specific species in a large sample of objects, e.g., SiO (Sch{\"o}ier et al. 2006). Just above the photosphere, the gas experiences the passage of shocks triggered by the pulsation of the star and these shocked gas layers are the locus of dust formation (Bowen 1988). These dense molecular gas layers has been detected by both observations with ISO (Tsuji et al. 1997) and near-infrared (IR) interferometry (Perrin et al. 2004). A semi-analytical model for the physics of these shocked regions based on the work of Fox \& Wood (1985) and Bertschinger \& Chevalier (1985) was proposed by Cherchneff (1996) assuming that the shock energy was dissipated in the immediate postshock gas by the collisional dissociation of H$_2$ and that subsequent cooling was provided by adiabatic expansion. Using a similar formalism for the inner wind of IRC+10216, Willacy \& Cherchneff (1998) modelled the gas-phase chemistry and showed that the collisional destruction of CO in the shocks could release free atomic oxygen and trigger the formation of OH and SiO. The formations of HCN, CO$_2$, and CS were later described in the inner wind of the O-rich Mira star IK Tau by Duari et al. (1999). These two studies therefore highlighted the importance of the shock-induced non-equilibirum chemistry in the inner wind of AGB stars in unleashing the synthesis of molecules that were not expected to form under photospheric TE conditions. More generally, Cherchneff (2006) studied the inner wind composition as a function of the carbon-to-oxygen (C/O) ratio of the stellar photosphere and confirmed the formation of O-bearing species in carbon stars and that of C-bearing molecules in O-rich AGB stars as a result of non-equilibrium chemistry induced by periodic shocks. Several observations of high energy rotational transitions of HCN, SiS, CS, and SiO in AGBs and supergiants at mm and submm wavelengths corroborated these results (e.g., Sch{\"o}ier et al. 2006, 2007,  Ziurys et al. 2007, 2009, Decin et al. 2008). 

A next step in understanding the complexity of the wind chemistry close to the stellar photosphere was achieved with Herschel and the confirmation of the widespread presence of water in the dust formation zone of several carbon stars (Decin et al. 2010b, Neufeld et al. 2010, 2011a, 2011b). Ag{\'u}ndez et al. (2011) proposed that the partial dissociation of $^{13}$CO and SiO by UV photons penetrating deep inside the clumpy stellar wind could produce free atomic oxygen and lead to the subsequent formation of H$_2$O. This scenario would destroy other species in the deep layers and have some impact on the water isotopologue abundances as discussed by Neufeld et al. (2011a). Both effects have not yet been tested observationally. Updating the inner wind model for IRC+10216, Cherchneff (2011a) showed that the non-equilibrium chemistry triggered by the shocks could form water in the dust formation zone in competition with the synthesis of SiO, with abundances that agree excellently with those derived from observations. 

In this study, we present a complete, updated chemical model of the inner wind of the carbon star IRC+10216, based on the non-equilibrium chemistry approach mentioned previously. The synthesis of classical molecules including water and several new species is considered in the gas-phase, comprising hydrides, chlorine and phosphorous compounds, hydrocarbons and aromatics, and gas-phase precursors to silicon carbide and metal-sulphide dust. The goal of such a study is to confirm the formation of already detected species and predict new molecules potentially observable by means of their high excitation transitions in the inner wind. In Section \ref{phys}, we give a brief description of the physics of the shocked layers, while the chemistry is discussed in Section \ref{chem}, the results for the various families of species are presented in Section \ref{res}, and a discussion follows in Section \ref{dis}.  

%__________________________________________________________________

\section{Physics of the inner wind}
\label{phys}

%---------------
\begin{table}
\caption{Stellar parameters for IRC+10216}             % title of Table
\label{tab1}      % is used to refer this table in the text
\centering                          % used for centering table
\begin{tabular}{l cl}        % centered columns (4 columns)
\hline\hline                 % inserts double horizontal lines
M$_{\star}$ (\Ms) & 2 &  1\\   
\rstar (cm) & 6.5$\times 10^{13}$& 2 \\
C/O  & 1.4 & 3 \\  
P (days) & 650 & 4 \\
r$_s$ (\rstar) & 1.2 & 2 \\
V$_s$ (km s$^{-1}$) & 20 & 2 \\
n$_{gas}$(r$_s$) (cm$^{-3}$) & 3.6 $\times 10^{13}$ & 5 \\
T$_{gas}$(r$_s$) (K) & 2062 & 5 \\
\hline                                   %inserts single line
\end{tabular}
\tablebib{(1)~Willacy \& Cherchneff 1998;
(2)~Ridgway \& Keady 1988; (3)~Winters et al. 1994; (4)~Witteborn et al. 1980;
(5)~Cherchneff 2011a.
}
\end{table}
%
%------

The stellar parameters used in this study for IRC+10216 are listed in Table \ref{tab1}. The stellar photosphere and the outer atmospheric layers above it are assumed at TE with a solar elemental composition (Asplund et al. 2009) except for carbon where the C/O ratio was set to 1.4 (Winters et al. 1994). The shocks are assumed to form at a radius r$_s=1.2$ \rstar~with a velocity V$_s$. The TE calculations are run until r$_s$, which is characterised by the temperature and number density listed in Table \ref{tab1}, and the TE abundances are used to characterise the unshocked gas. Since the pre-shock gas at r$_s$ has a high molecular component, the impact of the passage of periodic shocks is modelled by considering a post-shock gas that cools via both 1) the collision-induced dissociation of H$_2$ and 2) adiabatic expansion. The shock jump in density, temperature, and velocity is described by the Rankine-Hugoniot jump conditions applied to the pre-shock gas at r$_s$ for a shock velocity V$_s= 20~$km s$^{-1}$ (Ridgway \& Keady 1981). According to Fox \& Wood (1985), radiative processes do not operate in cooling the post-shock gas owing to the modest shock strength, for which the energy loss is provided by the endothermic dissociation of H$_2$ by collisions initiated by the prevalent chemical reaction
%-------
\begin{equation}
\label{h2diss}
{\rm H}_2 + {\rm H_2 \rightarrow H + H + H_2.}
\end{equation}
% --
Reaction \ref{h2diss} has a reaction rate given by k$_{diss} = 1.5 \times 10^{-9} \times \exp(-48346/{\rm T})$ cm$^3$ s$^{-1}$, where T is the gas temperature, and operates over a length l$_{diss}$ defined as

%-------
\begin{equation}
\label{h2len}
   {\rm l}_{diss} = \tau_{diss} \times {\rm v}_{gas} =  {\frac{1}{{\rm k}_{diss}\times {\rm n(H}_2{\rm)} }} \times \left[{\frac{{\rm V}_s}{{\rm N}_{jump}}}\right],
\end{equation}
%------
 where n(H$_2$) is the number density of H$_2$ in the post-shock gas, k$_{diss}$ is the rate of reaction \ref{h2diss}, V$_s$ is the shock velocity at r$_s$ as given in Table \ref{tab1}, and N$_{jump}$ is the Rankine-Hugoniot velocity shock jump. 
 
Once this cooling has proceeded over l$_{diss}$, we assume that later cooling occurs via adiabatic expansion. As modelled by Bertschinger \& Chevalier (1985), the fluid equations that describe the conservation of mass, momentum, and energy are parametrised and solved for the boundary conditions imposed by stellar gravity and the return of the shocked gas to its initial pre-shock position. Typical excursions of the gas layers over several pulsation periods are illustrated in Willacy \& Cherchneff (1998) and Cherchneff (2011a) for IRC+10216. The pre-shock gas temperature and density profiles are derived as a function of radius using the formalism of Cherchneff et al. (1992), where the impact of shocks on the gas density is described by an extended scale-height formalism for the initial conditions listed in Table \ref{tab1}. The derived pre- and post-shock gas parameters at various radius are those assumed in Cherchneff (2011a) and listed in Table \ref{tab2}.
  
%----------
\begin{table*}
\caption{Gas parameters in the dust formation zone of IRC+10216 taken from Cherchneff (2011a). The shocks form at r$_s =1.2$ \rstar. For each radius, the temperature and number density are given for the pre-shock gas, the gas after the shock front (in the collision-induced H$_2$ dissociation zone), and the gas at the beginning and the end of the adiabatic cooling zone. The temperature T is in Kelvin and the gas number density n$_{gas}$ in cm$^{-3}$.}             
% title of Table
\label{tab2}      % is used to refer this table in the text
\centering                          % used for centering table
\begin{tabular}{c c c c c c c c c c  }        % centered columns (4 columns)
\hline\hline                 % inserts double horizontal lines
r & Shock strength &\multicolumn{2}{c}{Preshock gas} & \multicolumn{2}{c}{Shock Front} & \multicolumn{2}{c}{Adiabatic expansion - start} & \multicolumn{2}{c}{Adiabatic expansion - end} \\    % table heading 
\hline                        % inserts single horizontal line
 (\rstar) & (km s$^{-1})$&T & n$_{gas}$ & T & n$_{gas}$ & T  & n$_{gas}$& T  & n$_{gas}$  \\
 \hline
   1.2 & 20.0 &2062 & 3.63(13)& 19725 &1.98(14) & 4409& 5.97(14)&1480 & 3.63(13)\\     % inserting body of the table
   1.5 & 17.9 &1803 & 8.24(12)& 15922 &4.40(13) & 3870& 1.29(14)&1290 & 8.24(12)\\ 
   2 & 15.5 &1517 & 1.44(12)& 12081 &7.59(12) & 3200& 2.14(13)&1080 & 1.44(12)\\ 
   2.5 & 13.9 &1327& 4.24(11)& 9779&2.21(12) &2750& 6.08(12)&951& 4.24(11)\\ 
   3& 12.6& 1190 & 1.69(11)& 8245 &8.73(11) & 2430& 2.35(12)&848 & 1.69(11)\\ 
   4& 11.0 &1001 & 4.51(10)& 6284 &2.29(11) & 1790& 4.48(11)& 711& 4.51(10)\\ 
   5 & 9.8&876 & 1.79(10)&5096 &8.94(10)&1550&1.71(11)&621 &1.79(10)\\
\hline
\end{tabular}                        %inserts single line
\end{table*}
%--------
%-----
\begin{table*}
\caption{Chemical species included in the inner wind model of IRC+10216.}             % title of Table
\label{tab3}      % is used to refer this table in the text
\centering                          % used for centering table
\begin{tabular}{l c c c c c c c}        % centered columns (4 columns)
\hline\hline                 % inserts double horizontal lines
Classical & H$_2$ & CO & SiO & HCN & C$_2$H$_2$  & SiS &  \\
\hline
Hydrocarbons &C$_2$H$_3$ & C$_3$H$_2$ & C$_3$H$_3$& C$_4$H$_2$ & C$_4$H$_3$ &C$_4$H$_4$ \\
\hline
Aromatics & C$_6$H$_5$& C$_6$H$_6$ & &  & & &  \\
\hline
C & CH & CH$_2$ &CH$_3$& C$_2$H & C$_2$ & C$_3$   & \\
\hline
O & OH & H$_2$O & O$_2$ &CO$_2$&HCO &   &   \\ 
\hline
N & NH & N$_2$ & CN &NO & NH$_2$& NH$_3$ &  \\
\hline
P & PH &P$_2$ & HCP& CP& PN& & \\
\hline
Si & SiN & SiC & SiC$_2$ & Si$_2$C$_2$ & & & \\
\hline 
S & CS  &SH & H$_2$S & SO &   & & \\
\hline
Mg & MgH &MgS & Mg$_2$&MgO & & & \\
\hline
Fe &FeH  & FeS& Fe$_2$&FeO & & & \\ 
\hline
F& HF & F$_2$   &ClF &  & & & \\ 
\hline 
Al & AlH & AlCl &   & & & & \\ 
\hline
Na &  NaH& NaCl &   & & & & \\ 
\hline
Cl & HCl &Cl$_2$ &  &  && & \\
\hline
K & KH& KCl &   & & & & \\ 
\hline\hline     
\end{tabular}
\end{table*}
%----

\section{Chemistry of the inner wind}
\label{chem}

New families of chemical species are considered along with the classical molecules already studied in previous models (Cherchneff 2006, 2011a). They include hydrides, halogens, in particular chlorine- and fluorine-bearing species, and finally phosphorous-bearing species. Small gas-phase molecular precursors to silicon carbide and metal-sulphide grains are also included. All species are listed in Table \ref{tab3} where the largest molecule is the aromatic ring of benzene, \ben. For metal hydrides, there is no available information on chemical reaction rates for the formation processes of several of them, namely FeH, MgH, NaH, KH, and PH, at the high temperatures characterising the inner wind. We thus assume rates similar to those of documented reactions using the principle of isovalence as a guideline. Generally, the typical rates of the dominant formation pathways that consists of the reaction of the metal with H$_2$ have an Arrhenius factor ranging from 10$^{-12}$ and 10$^{-9}$ cm$^{3}$s$^{-1}$, an activation energy barrier of a few 1000 K, and a mild temperature dependance (e..g, Cohen \& Westberg 1983). 

The chemistry of halogens including chlorine- and fluorine-bearing species is rather well-documented at high and intermediate temperatures. However, this is not the case for the phosphorous chemistry. Owing to the isovalence of phosphorous, P, with nitrogen, N, we assume for the P-bearing species chemical processes similar to those involving N and for which the rates are measured or calculated. Finally, the chemistry of silicon, Si, and sulphur, S, is poorly known and studied. For these elements, we restrict the chemical processes  to the set of reactions for which reaction rates are documented. 

All chemical pathways that lead to the formation of linear molecules, carbon chains, and aromatic rings, include neutral-neutral processes such as termolecular, bimolecular, and radiative association reactions, whereas destruction is described  by thermal fragmentation and neutral-neutral processes (i.e., oxidation reactions of hydrocarbons and all reverse processes of the formation reactions). No ions are considered in this chemistry because the UV stellar radiation field of IRC+10216 is too weak to foster the efficient photodissociation and ionisation of molecules. In total, 63 species were considered for a chemical network of 377 reactions. Details of all these processes are  provided in the online Appendix 1 available at the CDS, which gives the reactions included in the chemical network and their reaction rates. It contains the following information:
Column 1 lists the reaction number, Column 2 gives the reactants and  the products, Column 3 lists the A coefficient, Column 4 lists the n factor, Column 5 lists the activation energy in Kelvin, and Column 6 gives the reference for the data. The major differences between the present chemical network and that used in previous studies (e.g., Willacy \& Cherchneff 1998, Cherchneff 2006) are explained in detail in the Appendix of Cherchneff (2011a). In terms of formalism, three major changes are implemented. Firstly, the treatment of the reverse reaction of a specific process is changed. Several new rates have been measured in combustion and aerosol chemistry, and are now available. The calculation of the reverse rate from the equilibrium constant assuming detailed balance (i.e., Equation 4 in Willacy \& Cherchneff 1998) can lead to erroneous values when the gas temperature and density decrease. Therefore, we prefer to directly enter  the available measured or calculated rate values in the chemical network. When the information is unavailable, we make 'educated' guesses depending on the type of the reaction and its endo-(exo)thermicity. This approach allows us to test a variety of reaction paths that would be closed if one were to strictly apply detailed balance considerations. Secondly, the chemistry involving atomic silicon, Si, and Si-bearing species is restricted to reactions for which rates have been measured or calculated. Thirdly, as previously mentioned, the chemistry now describes the formation of a larger set of chemical species. This chemical scheme is used to solve a set of 63 stiff, coupled, ordinary differential equations (ODEs) at each radius of the inner wind. These coupled ODES are integrated over space (for the H$_2$ cooling region defined by Equation \ref{h2len}) and time (for the adiabatic expansion over a pulsation period) for the radii and corresponding shock strengths of Table \ref{tab2}. The post-shock abundances of the species at the end of the pulsation period and radius r$_i$ are used as pre-shock initial abundances for the successive radius r$_{i+1}$. 

In all, the chemical network aims to provide a comprehensive and coherent chemical description of the wind, where known trends are reproduced or new trends are presented. It does not aim to give an exact description of all chemical processes in the wind, which is an impossible task to perform. Owing to the uncertainties in both our theoretical description of the inner wind and observational data analysis, we consider in the following sections that a good agreement has been reached when modelled abundances and derived values from observations differ by at most a factor of ten. 

\section{Results}
\label{res}

%------
\begin{table*}
\caption{Predicted abundances (with respect to H$_2$) of selected species. TE abundances are given at r$_s$ and modelled abundances are listed for r$_s$ and 5 \rstar. Most recent abundances (with respect to H$_2$) derived from observations are also listed when available with the corresponding reference. '$N$' corresponds to abundances lower than $ 1\times 10^{-15}$. }             
% title of Table
\label{tab4}     
\centering                          % used for centering table
\begin{tabular}{cccc cl }       
\hline\hline                
 % inserts single horizontal line
 Species & TE at $r_s$& Predicted at r$_s$& Predicted at 5 \rstar&Observed&  Reference \\
 \hline
 CO  & $1.5 \times 10^{-3}$&$8.1 \times 10^{-4}$&$9.1 \times 10^{-4}$&$1.0 \times 10^{-3}$ & Herschel $-$ Cernicharo et al. (2010a) \\   
 HCN & $4.3 \times 10^{-5}$& $3.2 \times 10^{-5}$&$5.3 \times 10^{-6}$&$8.0 \times 10^{-6} - 2.0 \times 10^{-5}$& IRAM $-$ Cernicharo et al. (2011) \\
 C$_2$H$_2$ & $9.2\times 10^{-5}$& $2.0 \times 10^{-4}$&$1.5 \times 10^{-4}$&$8.5 \times 10^{-5}$& TEXES/IRTF $-$ Fonfr{\'i}a et al. (2008) \\
 N$_2$ & $7.5 \times 10^{-5}$ &$4.4 \times 10^{-5}$&  $5.8 \times 10^{-5}$ & & \\
 SiS & $4.3 \times 10^{-6}$ & $2.2 \times 10^{-5}$&$2.1 \times 10^{-5}$&$2.0 \times 10^{-6} - 8.0 \times 10^{-6}$ & PACS/SPIRE Herschel $-$ Decin et al. (2010a) \\
 CS & $3.6 \times 10^{-5}$& $5.6 \times 10^{-6}$&$6.4 \times 10^{-6}$&$9.3 \times 10^{-6}$& SMA - Patel et al. (2009) \\
 SiO & $4.1 \times 10^{-9}$& $8.4 \times 10^{-6}$&$3.6 \times 10^{-8}$ & $2.0 \times 10^{-8} - 3.0 \times 10^{-7}$& PACS/SPIRE Herschel $-$ Decin et al. (2010a) \\
 H$_2$O & $2.0\times 10^{-12}$& $9.9 \times 10^{-5}$&$7.1 \times 10^{-7}$ & $8.0 \times 10^{-8}$ & HIFI/Herschel $-$ Neufeld et al. (2011a) \\
     & --    &--& --&$ > 1 \times 10^{-7}$\tablefootmark{a} & PACS/SPIRE Herschel $-$ Decin et al. (2010b) \\
SiC$_2$ & $4 \times 10^{-7}$& $3.8 \times 10^{-11}$& $5.5 \times 10^{-8}$&$2.0 \times 10^{-7}$ & HIFI/Herschel $-$ Cernicharo et al. (2010a)\\
HCl & $5.4 \times 10^{-7}$&$6.3 \times 10^{-7}$& $4.9 \times 10^{-7}$ &$1.0 \times 10^{-7}$ & HIFI/Herschel $-$ Ag{\'u}ndez et al. (2011)\\
HF & $1.2 \times 10^{-7}$&$7.3 \times 10^{-8}$& $7.2 \times 10^{-8}$ & $8.0 \times 10^{-9}$ & HIFI/Herschel $-$ Ag{\'u}ndez et al. (2011)\\
NaCl & $2.4\times 10^{-12}$&$6.7\times 10^{-9}$& $1.2 \times 10^{-8}$ &$2 \times 10^{-9}$ & SMT/ARO $-$ Milam et al. (2007)\\
AlCl & $1.6 \times 10^{-10}$&$1.8 \times 10^{-10}$& $1.3 \times 10^{-7}$&  $3.5 \times 10^{-8}$& IRAM $-$ Ag{\'u}ndez (2009) \\
KCl & $6.4 \times 10^{-13}$& $7.2 \times 10^{-10}$&$1.2 \times 10^{-9}$ & $7.0 \times 10^{-10}$& IRAM $-$ Ag{\'u}ndez (2009) \\
AlH & $3.8 \times 10^{-9}$&$4.6 \times 10^{-6}$&  $2.0\times 10^{-8}$&  $< 1.6 \times 10^{-6}$\tablefootmark{c}& PACS/SPIRE Herschel $-$ Cernicharo et al. (2010b)\\
NaH & $1.1 \times 10^{-11}$& $1.4 \times 10^{-6}$&$9.9\times 10^{-12}$ &$< 3.0\times 10^{-9}$\tablefootmark{c} &PACS/SPIRE Herschel $-$ Cernicharo et al. (2010b) \\
KH & $2.4 \times 10^{-12}$& $1.1 \times 10^{-7}$&$9.0\times 10^{-13}$ & $< 3.0\times 10^{-9}$\tablefootmark{c} &PACS/SPIRE Herschel $-$ Cernicharo et al. (2010b)\\
MgH & $3.7 \times 10^{-9}$& $7.9 \times 10^{-9}$&$5.2\times 10^{-12}$ & $< 4.0\times 10^{-8}$\tablefootmark{c}& PACS/SPIRE Herschel $-$ Cernicharo et al. (2010b)\\
FeH &$1.4 \times 10^{-10}$ &$6.5 \times 10^{-9}$& $4.3\times 10^{-12}$ & $ < 1.0\times 10^{-9}$\tablefootmark{c}& PACS/SPIRE Herschel $-$ Cernicharo et al. (2010b)\\
PH & $1.8 \times 10^{-9}$& $5.6 \times 10^{-11}$&$N$& & \\
SH & $1.4 \times 10^{-7}$&$2.3 \times 10^{-12}$&$ N $& & \\
NH & $2.8\times 10^{-11}$& $2.6 \times 10^{-13}$&$N$ & & \\
HCP &$2.6\times 10^{-7}$ & $1.0\times 10^{-8}$& $8.2 \times 10^{-9}$& $1.4 \times 10^{-8}$\tablefootmark{b} & IRAM $-$ Ag{\'u}ndez et al. (2007)\\
 &--&-- & --&  $3.0\times 10^{-8}$& SMT/ARO $-$ Milam et al. (2008) \\
PN & $1.0\times 10^{-10}$& $4.4 \times 10^{-7}$&$4.3\times 10^{-7}$ &$3.0 \times 10^{-10}$ & SMT/ARO $-$ Milam et al. (2007) \\
P$_2$& $3.0 \times 10^{-12}$& $1.3 \times 10^{-13}$&$2.8\times 10^{-9}$ & & \\ 
FeS& $1.2 \times 10^{-12}$& $1.2 \times 10^{-10}$&$2.2\times 10^{-9}$ & & \\ 
MgS& $6.2\times 10^{-13}$&$1.5 \times 10^{-10}$& $2.7\times 10^{-9}$ & & \\ 
\hline
\end{tabular}
\tablefoot{
\tablefoottext{a}{Refer to the maximum value reached by H$_2$O abundance distribution with respect to H$_2$, }
\tablefoottext{b}{Refer to the HCP abundance for r$> 20$ \rstar, }
\tablefoottext{c}{3 $\sigma$ upper limits to abundances. }
} 
\end{table*}
%--------

\begin{figure}
   \centering
   \includegraphics[angle=0,width=8.9cm]{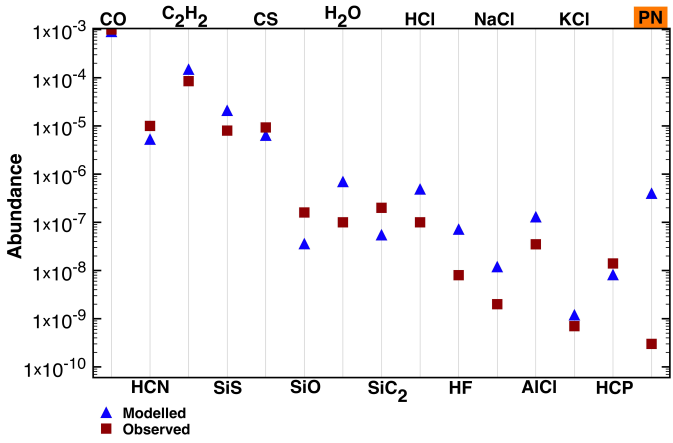}
      \caption{Observed and modelled abundances (with respect to H$_2$) for most of the molecules listed in Table \ref{tab4}. Abundances are at phase $\theta = 1$ and 5 \rstar. The agreement between modelled and observed values is satisfactory and less than a factor of ten for all species apart from PN (for CO both values overlap).
              }
         \label{fig0}
   \end{figure}

%__________________________________
Results for the dominant chemical species in the inner wind are summarised in Table \ref{tab4}. We list the abundance values derived by applying TE to the gas conditions met at r$_s$, as well as the non-equilibrium abundances at r$_s$ and 5 \rstar, and abundances derived from the most recent observations. As stated before, all modelled and observational values differ by at most a factor of ten, except for PN (see \S \ref{phos}). This good agreement is illustrated in Figure \ref {fig0} where we compare the modelled and observed abundances with respect to H$_2$ at 5 \rstar. Discrepancies between values derived from TE and non-equilibirum chemistry for O-bearing species are clear from inspection of Table \ref{tab4}. These discrepancies highlight once more the importance of the shock chemistry in the dust formation zone of AGB stars. We discuss below the results for the specific chemical families under study. 

%-------------------------------------------------------------
%--------------

\subsection{Prevalent molecules}
\label{prev}

%-------------------------------------------------------------
   \begin{figure}
   \centering
   \includegraphics[angle=0,width=8.9cm]{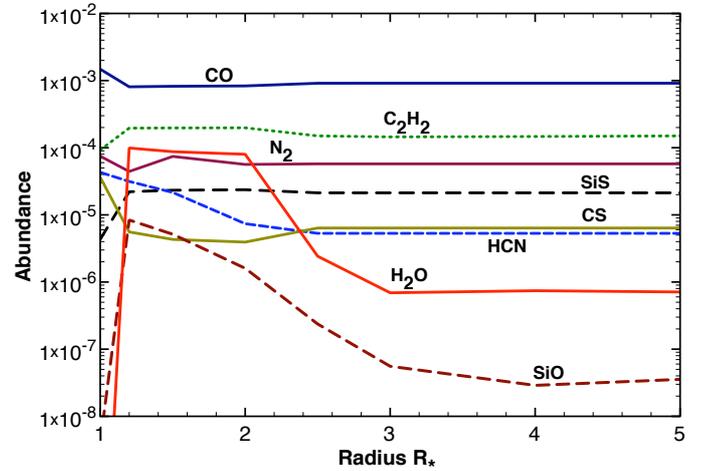}
      \caption{Abundances (with respect to H$_2$) of the prevalent molecules formed in the inner wind as a function of radius. Abundances are those at phase $\theta = 1$, i.e., after one full oscillation cycle. Abundances derived by assuming thermodynamic equilibrium at r$_s$ are moved to radius 1 \rstar~for clarity.
              }
         \label{fig1}
   \end{figure}
%
%__________________________________

%-------------------------------------------------------------
   \begin{figure}
   \centering
   \includegraphics[angle=0,width=8.9cm]{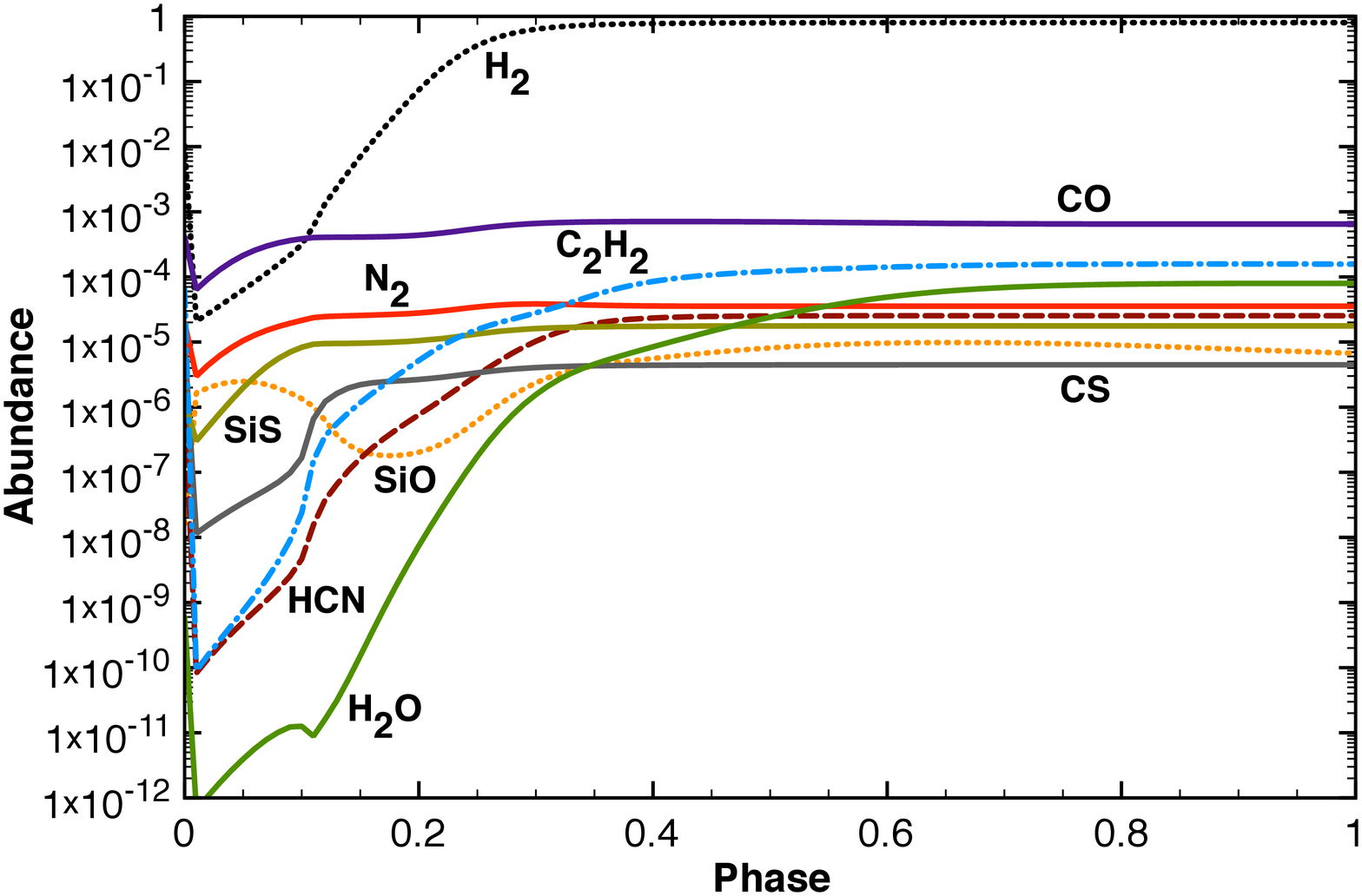}
      \caption{Abundances (with respect to total gas) of the prevalent molecules shown in Figure \ref{fig1} as a function of phase $\theta$ at the shock formation radius r$_s$. Destruction occurs at small values of $\theta$ but the molecules reform as the gas cools down.            }
         \label{fig2}
   \end{figure}
%
%__________________________________

A group of species dominates the molecular phase of the shocked, inner wind along with H$_2$. It consists of CO, HCN, CS, N$_2$, C$_2$H$_2$, SiS, SiO, and H$_2$O, whose abundances relative to H$_2$ are illustrated as a function of radius in Figure~\ref{fig1}, while their abundances with respect to total gas and as a function of pulsation phase $\theta$ at the shock formation radius r$_s$ are shown in Figure~\ref{fig2}. All molecules experience destruction in the post-shock gas at  $\theta < 0.05$ as seen in Figure~\ref{fig2}. The destruction is more or less severe depending on the species and no large discrepancies may exist between TE and non-TE abundances for specific molecules at the end of one oscillation. However, TE abundance values for several species (CS, SiS, SiO, and H$_2$O) differ drastically from those obtained from shock-induced, non-equilibrium chemistry, as already stressed in existing studies (Willacy \& Cherchneff 1998, Cherchneff 2006, 2011a). 

The formation of O-bearing species, namely H$_2$O and SiO, results from the collision dissociation of CO in the post-shock gas. For the specific conditions of the IRC+10216 model, between 10 \% and 20 \% of CO molecules are destroyed at r$_s$ in the H$_2$ dissociation cooling part, while they quickly reform in the adiabatic expansion at phase $\theta > 0.05$, making CO the main provider of atomic oxygen in the post-shock gas. The formation of H$_2$O is thus triggered by the reactions (Cherchneff 2011a)
% ---
\begin{equation}
\label{wat1}
   {\rm O + H_2} \rightarrow {\rm OH + H}
\end{equation}
and 
\begin{equation}
\label{wat2}
   {\rm OH + H_2} \rightarrow {\rm H_2O + H}
\end{equation}
The reaction given in Eq. \ref{wat2} is in competition with the formation of SiO via the reaction 
%----
\begin{equation}
\label{sio}
   {\rm Si + OH} \rightarrow {\rm SiO + H}.
\end{equation}
% ---
The rate of the backward process of the reaction in Eq. \ref{sio} is unknown but the reaction has an endothermicity of $\sim$ 40000 K at 2000K.  As mentioned in Section \ref{chem}, we assume a low rate for that reaction (k $= 1 \times 10^{-15}$ cm$^{3}$ s$^{-1}$) that reflects its low efficiency over the temperature range of interest. We considered the resulting water abundance for different rate values and temperature dependences. The water abundance always shows a trend similar to that reported here, i.e, a high inner value that decreases at radius $r > 2$ \rstar~to reach a typical value of $\sim 1\times10^{-7}$. In contrast, the rate of the backward reaction in Eq. \ref{sio} was explicitly calculated from the equilibrium constant given in the study of Willacy \& Cherchneff (1998). This previous rate had a very low value for temperatures lower than 2000K, and contributed in part to the non-replenishment of OH and the disappearance of H$_2$O at larger stellar radii.

The prevalent molecules in Figure~\ref{fig1} come from different chemical families that are chemically linked together by the indirect key-role of the over-abundant H$_2$ species. Molecular hydrogen is primarily involved in the formation processes of both various members of chemical families (e.g., hydrides, hydrocarbons) and specific molecules such as OH and H$_2$O. The destruction or formation of H$_2$ can then impact all chemical families, which become interrelated. This is particularly true for water, owing to its link to the hydrocarbon family. The destruction of hydrocarbons, starting with \ace, indeed releases H$_2$. Through the synthesis of hydroxyl, OH, via the reaction of atomic O with H$_2$ in the post-shock gas, H$_2$O is thus linked to the hydrocarbon family. As seen above, water is also linked to the Si-bearing species as both H$_2$O and SiO  species are competitors in the depletion of OH. Atomic Si is also efficiently included in silicon monosulphide, SiS, and in silicon carbide, SiC, in the hot post-shock gas at r$_s$. Therefore, the Si-bearing species through their connection to SiO impact the water abundance, whose value listed in Table \ref{tab4} is slightly higher than the value derived by Cherchneff (2011a).  The chemistry of SiC dust precursors has been extended to the first ring (SiC)$_2$ in the present study, and as discussed in \S~\ref{other}, the formation of SiC and the rings SiC$_2$ and (SiC)$_2$ proceeds very early on. The more atomic Si is trapped in SiC, the less Si is able to react with OH to form SiO. Thus, the SiO abundance listed in Table \ref{tab4}~is lower than that derived by Cherchneff (2011a) by a factor of about two, resulting in a higher water abundance. However, the chemical trends and processes in both studies are alike. The formation pathways to other important molecules such as \ace~, CS, and HCN were discussed in detail by Cherchneff (2006) for a carbon star with a C/O ratio equal to 1.1 and similar chemical routes operate in the inner wind of IRC+10216. Overall and as seen from Figure \ref{fig0}, the most abundant species have modelled abundances that agree very well with values derived from observations.

\subsection{Hydrides}

\label{hydri}
%-------------------------------------------------------------
   \begin{figure}
   \centering
   \includegraphics[angle=0,width=8.9cm]{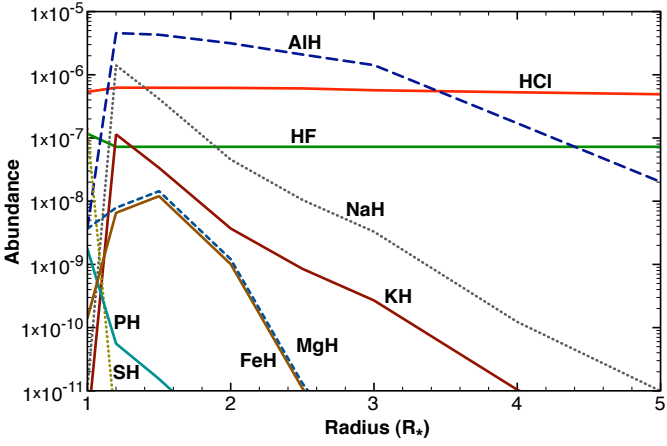}
      \caption{Abundances (with respect to H$_2$) of the hydride species formed in the inner wind as a function of radius. Abundances derived by assuming thermodynamic equilibrium at r$_s$ are moved to radius 1 \rstar~for clarity.  }
         \label{fig3}
   \end{figure}
%
%__________________________________

The interest of studying light hydrides was rekindled with the launch of Herschel. The rotational spectra of these species lie in the submm and far-IR domains and are difficult to observe from Earth. The detection by HIFI onboard Herschel of the J=1$\rightarrow$0,  2$\rightarrow$1 and 3$\rightarrow$2 transitions of HCl and the J=1$\rightarrow$0 transition of HF in IRC+10216 was reported by Ag{\' u}ndez et al. (2011). 

The gas-phase chemistry of many light hydrides has never been clearly characterised and only a few measured rates have been documented. However, their production in the laboratory from the gas phase occurs via the reaction of a metal vapour with hydrogen (Ozin \& McCaffrey 1984).  For our model, we consider all known chemical reactions and extend the documented chemistry to some species according to the isovalence of specific elements (e.g., N and P). Following the prescriptions of experimental studies, we model the formation of hydrides according to 
%-------
\begin{equation}
\label{hyd1}
   {\rm X + H}_2 \rightarrow {\rm XH + H},
\end{equation}
%------
where X represents any atomic species. H$_2$ is destroyed in the strongest shocks up to $\sim$ 3 \rstar~but efficiently reforms in the post-shock gas at phase $\theta > 0.3$ (see Figure \ref{fig2}). The large H$_2$ reservoir in the gas layers insures that most hydrides are formed following the reaction in Eq. \ref{hyd1}. The abundance variation with radius of the major hydrides is illustrated in Figure \ref{fig3}. Hydrides have variations that depend on the species. The abundances of AlH are high out to 3 \rstar~but decrease at larger radii owing to the formation of aluminium chlorine, AlCl (see \S \ref{halo} below). A similar behaviour applies to NaH and KH with the formation of NaCl and KCl. In contrast, HCl and HF have consistently high abundances ($x({\rm HCl}) \sim 4\times 10^{-7}$ and $x({\rm HF}) \sim 7\times 10^{-8}$) in the inner wind while MgH, FeH, PH, and SH have extremely low abundances that are well below the PACS/SPIRE $3\sigma$ detection limits estimated by Cernicharo et al. (2010b). 

The Herschel detection of the J=1$\rightarrow$0 transition of HF by Ag{\' u}ndez et al. (2011) points to a constant abundance of $\sim 8 \times 10^{-9}$ with respect to H$_2$ extending from the inner envelope up to 45 \rstar. This value is lower by a factor of nine than that derived from TE in the photosphere, where most of the fluorine is in the form of HF. To reconcile these two values, they claim that F must be depleted on dust grains in the inner wind. Our calculated HF abundance in Table \ref{tab4}~has the  constant value of $7.25 \times 10^{-8}$, corresponding to the solar abundance for F. This value reflects the quick conversion of fluorine into HF by its reaction with H$_2$ in the post-shock gas at all radii in the dust formation zone extending to 5 \rstar. The result confirms that HF acts as the main reservoir of fluorine in AGB stars. The discrepancy between our kinetic results and those of Ag{\' u}ndez et al. is difficult to quantify. Our abundance variation through the inner wind agrees well with the abundance profile that is required to reproduce the HF HIFI data, i.e., a constant abundance distribution extending up to 45 \rstar, but our calculated abundance value is higher. Ag{\' u}ndez et al. quote that the error in the radiative transfer model is a factor of two, while the reaction rate for the gas-phase formation of HF from H$_2$ has been measured and is well-documented. The use of other available rate values does not change the present result, which indicates that fluorine is totally depleted into hydrogen fluoride in the dust formation zone.  
 
%----
\subsection{Chlorines}
\label{halo}
%-------------------------------------------------------------
   \begin{figure}
   \centering
   \includegraphics[angle=0,width=8.9cm]{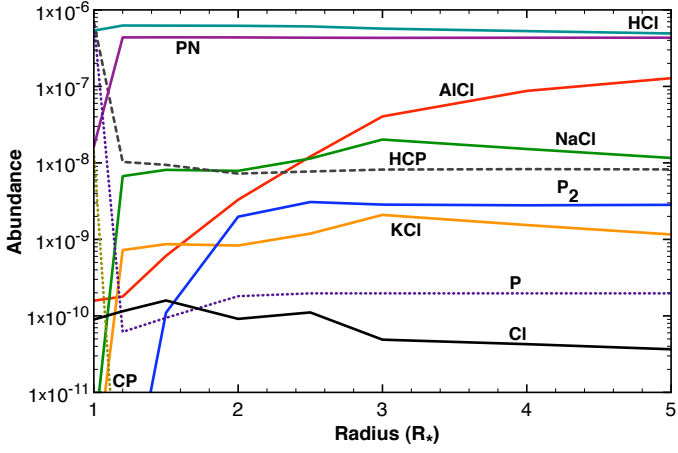}
      \caption{Abundances (with respect to H$_2$) of chlorine- and phosphorus-bearing species as a function of radius. Abundances derived by assuming thermodynamic equilibrium at r$_s$ are moved to radius 1 \rstar~for clarity.  }
         \label{fig4}
   \end{figure}
%
%__________________________________

Chlorine (Cl) has a solar abundance with respect to hydrogen of $\sim 3.2 \times 10^{-7}$ (Asplund et al. 2009), and is present at TE in the photosphere in the form of both HCl and Cl of comparable abundances with respect to H$_2$ of $\sim$ 2.6 $\times 10^{-7}$. Chlorine-bearing species have long been observed in circumstellar envelopes of AGB stars. Recent large molecular surveys confirm the presence of NaCl, AlCl, and KCL in IRC+10216 (Tenenbaum et al. 2010) when the hydride HCl was detected with Herschel and unambiguously found very close to the star (see \S~\ref{hydri}). Shinnaga et al. (2009) also identified KCl lines in their e-SMA observations that have a very compact distribution centred on the star. Results for the prevalent Cl-bearing species are shown in Figure \ref{fig4}. Clearly, HCl is still the dominant Cl-bearing species in the dust formation zone, and readily forms at r$_s$ with an almost constant abundance throughout the inner wind and a value of $3.9 \times 10^{-7}$ at 5 \rstar. This values perfectly agrees with that derived by Ag{\'u}ndez et al. (2011) and is very close to the TE abundance of HCl at r$_s$. However, HCl is destroyed in the shock at r$_s$ and reforms in the post-shock region at phase $\theta \ge 0.2$ to finally reach an abundance after one oscillation very close to the TE value. Other Cl-bearing species, namely AlCl,  NaCl, and KCl in decreasing order of importance, are also present in the wind and their production originates directly from that of HCl via the reaction
%-------
\begin{equation}
\label{chlo1}
{\rm   X + HCl} \rightarrow {\rm XCl + H},
\end{equation}
%------
where X is the metal. All documented reactions of Eq. \ref{chlo1} are fast, with activation energy barriers of a few thousand degrees (e.g., Husain \& Marshall 1986), requiring high temperatures to proceed. 

Very much like H$_2$ for hydrides, HCl acts as the production agent of Cl-bearing species in the dust formation zone of AGB stars. The chemistry of chlorine depends essentially on the hydrogen and the chlorine content of the gas through HCl, and is thus independent of the C/O ratio of the photosphere. Since HCl is a rather stable molecule (with a dissociation energy D$_0$ = 4.4 eV), it should also form efficiently in oxygen-rich AGB stars and by a chemical pathway similar to the reaction in Eq. \ref{hyd1} owing to the large H$_2$ reservoir of AGB winds. Other Cl-bearing species will also form via the reaction in Eq. \ref{chlo1} provided that metal atoms are available in the gas phase. HCl and other Cl-bearing species are thus expected to be present with intermediate-to-high abundances in the inner wind of oxygen-rich AGB stars. The detection of high energy transitions of NaCl towards the O-rich supergiant VY CMa and the O-rich Mira IK Tau by Milam et al. (2007) indeed indicates that these two species form close to the star, with abundances of 5$\times 10^{-9}$ and 4$\times 10^{-9}$, respectively. Our results corroborate these observations and we predict that HCl, NaCl, and KCl should be observable in the dust formation zone of O-rich evolved stars. The abundance of AlCl may be lower because 1) the Al-bearing species AlOH has a high abundance in the wind acceleration zone - Tenenbaum \& Ziurys (2010) derived a AlOH abundance of $\sim 1 \times 10^{-7}$ for VY Cma - and 2) a large fraction of Al is expected to be depleted in alumina, Al$_2$O$_3$, in the stellar wind. In reality, AlCl in VY CMa was not detected  by Tenenbaum \& Ziurys (2010).  
%---
\subsection{Phosphorous bearing species}
\label{phos}
The phosphorous-bearing molecules HCP, CP, C$_2$P, and PN have been detected in the wind of IRC+10216 (Gu{\'e}lin et al. 1990, Ag{\'u}ndez et al. 2007, Milam et al. 2008, Halfen et al. 2008, He et al. (2008), Tenenbaum et al. 2010). While the shapes of the C$_2$P and CP line profiles are indicative of a shell-like distribution with a formation locus in the outer envelope induced by UV photodissociation, HCP and PN have been found close to the star. The chemistry of phosphorous is poorly documented and we use the isovalence of P with N to estimate the rates of  a set of basic formation and destruction processes derived from the equivalent processes involving N. We consider a few molecules including PN, HCP, CP, and P$_2$. The latter species was chosen to reflect the refractory nature of phosphorous and its ability to form clusters. Results for P-bearing molecules are shown in Figure \ref{fig4}. PN is the prevalent phosphorous compound followed by HCP and P$_2$. PN is chiefly formed at r$_s$ by the two reactions
%-------
\begin{equation}
\label{pn1}
   {\rm N + CP} \rightarrow {\rm PN + C}
\end{equation}
%------
and 
%-------
\begin{equation}
\label{pn2}
   {\rm CN + CP} \rightarrow {\rm PN + C_2}
\end{equation}
%------
Similar reactions for the isovalent element N have measured rates at high temperatures. The resulting abundances for PN are quite high ($\sim 3 \times 10^{-7}$) through the inner wind with a rapid formation at r$_s$ where the molecule reaches its final abundance at phase $\theta = 0.5$ in the post-shock excursion. This is coherent with the fact that PN mimics N$_2$ which shows a similar behaviour: a rapid synthesis in the post-shock gas at r$_s$ and a constant high value across the inner wind (see Figure \ref{fig1}). At r$_s$ and at the early phases of the post-shock adiabatic excursion, phosphorous is quickly integrated into CP which later distributes P into HCP and PN. Milam et al. (2008) observed several rotational lines of PN and HCP in the inner wind of IRC+10216 and derived a low PN abundance of $\sim 3 \times 10^{-10}$. Our modelled value is greater by a factor of 1000. However, our derived HCP abundance of $8.2 \times 10^{-9}$ agrees well with the observations by  Ag{\'u}ndez et al. (2007) and Milam et al. (2008). The former study derived an abundance value of $1.4 \times 10^{-8}$ for radii larger than 20 \rstar, and claimed that a depletion onto dust grains is necessary to reconcile the high abundances derived assuming TE in the dust formation region with those at 20 \rstar. The present results indicate that HCP depletion in dust may not be necessary because the high TE abundances at r $<$ r$_s$ ($\sim 3.0 \times 10^{-7}$) quickly drop to $8.2\times 10^{-9}$ at larger radii owing to the non-equilibrium chemistry and the partial conversion of P into CP and PN in the post-shock gas.  

The high abundances of PN obtained in the model compared to observations force us to test the P chemistry and assess the conditions for which the conversion of CP into HCP and PN is effective. We decrease all chemical rates by a factor of ten but have no success in diminishing the PN abundance in the inner wind. We also lower the rates of the two forward reactions given in Eqs. \ref{pn1} and \ref{pn2}, by a factor of 100, which results in decreasing the PN abundance to $\sim 3.5 \times 10^{-8}$ but in increasing the HCP abundances to $3.2 \times 10^{-7}$, which is high to agree with observations. The assumption of isovalence between P and N requires a phosphorous chemistry in which HCP and PN mimic HCN and N$_2$. In the inner wind, nitrogen is distributed between these two species in almost equal amounts, a result corroborated by the excellent agreement of HCN abundances with values derived from observations. In the case of P, this distribution is not observed and the low PN abundance derived by Milam et al. (2008) indicates a low efficiency channel for the conversion of CP into PN in the dust formation zone. According to this result, they deduce that if PN/N$_2$ were approximately equal to P/N, the N$_2$ abundance would be very low ($\sim 1 \times 10^{-7}$). Such a low value contradicts the high N$_2$ abundance given by TE in the photosphere and fostered by the non-equilibrium chemistry of the inner wind (see Table \ref{tab4}). The discrepancy regarding the abundances of PN questions the validity of the assumption of isovalence in the case of phosphorous, highlights the different chemical processes that may control the phosphorous chemical family, and points to the need for more high resolution observations of high energy transitions of PN in the inner wind. 

Finally, inspection of Figure \ref{fig4} reveals the presence of a modest amount of phosphorous dimers, P$_2$, synthesized in the inner wind with x(P$_2$) $\sim 3\times 10^{-9}$ at 5 \rstar. As P has even lower abundances than P$_2$, a small population of P clusters may grow from P$_2$ coalescence but the aggregation process will terminate with the formation of the stable tetrahedral P$_4$ cluster, as observed in laser ablation of red phosphorous crystals (Bulgakov al. 2002). At most, a P$_4$ abundance of $\sim 1\times 10^{-10}$ may form with a left-over population of dimers with abundances $\sim 1 \times 10^{-9}$. Therefore, phosphorous clusters should not be prevalent condensates in the dust formation zone of carbon stars.  
%---
%\subsection{Other molecules of interest: H$_2$S, NH$_3$ and CH$_4$}

%Recent observations with the IRAM telescope have shown evidence for NH$_3$ transitions having a flat-top profile. See Cherchneff (2011). 
%---------------

\subsection{Carbon dust precursors: Hydrocarbons and aromatics}
\label{arom}

\begin{figure}
   \centering
   \includegraphics[angle=0,width=8.9cm]{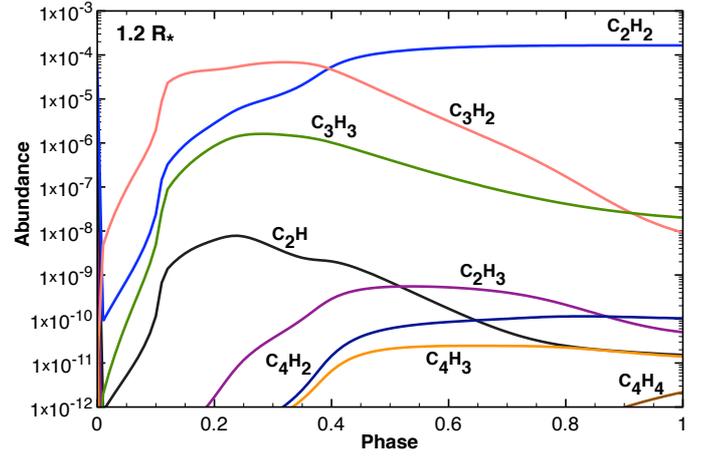}
      \caption{Abundances (with respect to H$_2$) of hydrocarbon species as a function of pulsation phase $\theta$ for the 20 km s$^{-1}$ at r$_s=1.2$ \rstar. No aromatic forms at these small radii owing to the presence of the oxidation agents H$_2$O and OH. 
              }
         \label{fig5}
   \end{figure}

\begin{figure}
   \centering
   \includegraphics[angle=0,width=8.9cm]{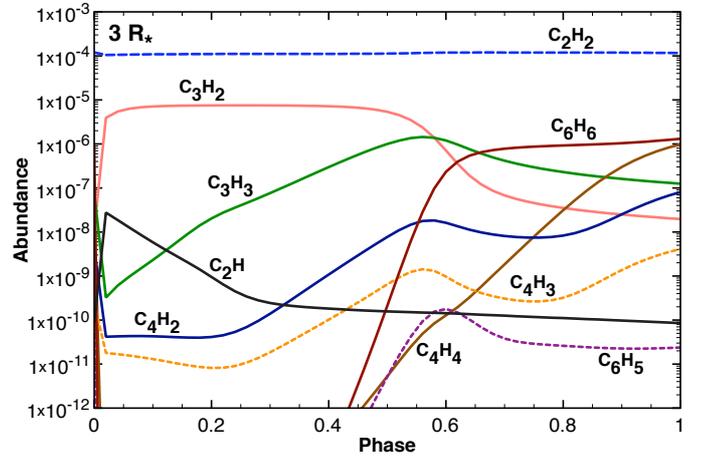}
      \caption{Abundances (with respect to H$_2$) of hydrocarbon and aromatic species as a function of pulsation phase $\theta$ for the 12.6 km s$^{-1}$ at 3 \rstar. Benzene starts forming at the low gas temperatures ($\sim 1000$ K) encountered at $\theta \geq 0.4$. 
              }
         \label{fig6}
   \end{figure}

\begin{figure}
   \centering
   \includegraphics[angle=0,width=8.9cm]{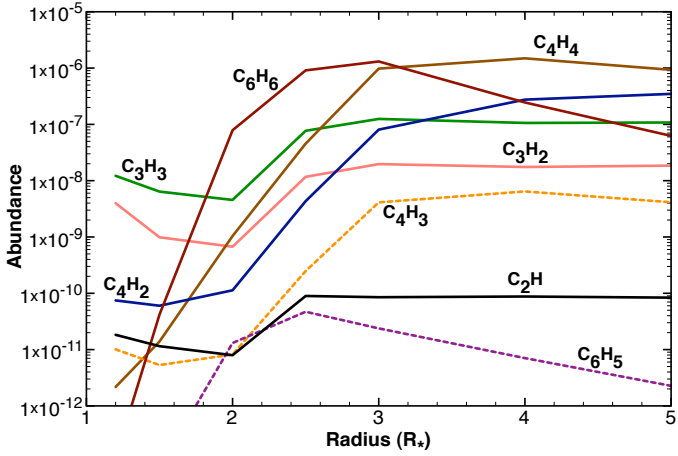}
      \caption{Abundances (with respect to H$_2$) of hydrocarbon and aromatic species as a function of radius, starting at r$_s=1.2$ \rstar. Benzene starts forming at $r > 2$ reaching a peak abundance value at 3 \rstar. Hydrocarbons such as C$_4$H$_4$, C$_4$H$_2$, and propargyl \prop~are also abundant in the dust formation zone.              }
         \label{fig6a}
   \end{figure}

The formation of the first aromatic ring of benzene represents a bottleneck to the formation of polycyclic aromatic hydrocarbon (PAH) species and their growth. In the chemical scheme, it is described by the recombination of two propargyl radicals, {\prop} which is the dominant closure pathway, and the reaction of  buten-3-ynyl radicals, C$_4$H$_3$, with \ace. These two routes are the prevalent channels to aromatic formation in sooting flames on Earth (Cherchneff 2011b). The formation of \prop~results from the reaction of \ace~with methylene, CH$_2$, in the immediate post-shock region at r$_s$. Abundances with respect to H$_2$ are shown in Figure \ref{fig5} for the 20 km s$^{-1}$ shock at r$_s$. At the high post-shock gas temperatures, only stable hydrocarbons such as \ace~, C$_3$H$_2$, and \prop~can form in large amounts. In Figure \ref{fig6}, the abundances for similar species are shown for a shock strength of 12.6 km s$^{-1}$ at 3 \rstar, with the appearance of \ben~at phases $>$ 0.4. As apparent in Figure \ref{fig1}, once \ace~forms at r$_s$, it stays abundant over the inner wind region, providing a large reservoir to grow hydrocarbons, specifically C$_3$H$_2$. Once \prop~starts to form from the reaction of C$_3$H$_2$ with H$_2$, \ben~quickly builds up when the gas temperature drops. Indeed, higher temperature values favour the reverse routes to the formation of  \ben, which are endothermic channels. The clincher to build up the \ben~ring are thus the lower gas temperatures encountered at phases $>$ 0.4. The abundances of hydrocarbons and aromatics as a function of radius are illustrated in Figure \ref{fig6a}. The formation of aromatics is delayed to r $\ge$ 2.5 \rstar~because oxygen-bearing species are present at smaller radii. Water and hydroxyl are the main oxidation agents to hydrocarbons and aromatics in the gas, and \ben~forms when the O-bearing species abundances drop at r $\geq$ 2.5 \rstar (see Figure \ref{fig1}). According to Figure \ref{fig6a}, the prevalent hydrocarbon species that escape the inner envelope are C$_4$H$_4$, C$_4$H$_2$, and \prop, although the growth and condensation of aromatics to amorphous carbon (AC) grains may alter this result to some extent.
%-------------------------------------------------------------
%
%__________________________________
%-------------------------------------------------------------
%
%__________________________________

%------
\begin{table*}
\caption{Derived dust-to-gas mass ratio as a function of radius in the dust formation zone of IRC+10216. We list the pre-shock gas number density,  acetylene and benzene abundances, the total number of \ace~and~\ben~molecules, the \ben~yield ($\equiv$ number of C atoms initially in \ace~locked into \ben), the mass of coronene \coro, the total number of \ace~molecules depleted in the growth from \ben~to~\coro, the corresponding number fraction of \ace~depleted in the growth from \ben~to \coro~in $ \%$, and the mass of AC dust formed (see text for detail). The region of efficient AC dust formation corresponds to numbers in boldface. }             
% title of Table
\label{tab5}      % is used to refer this table in the text
\centering                          % used for centering table
\begin{tabular}{l c c c c c c c }        % centered columns (4 columns)
\hline\hline                 % inserts double horizontal lines
 % inserts single horizontal line
 r (\rstar) & 1.2&1.5& 2 & {\bf 2.5} &{\bf 3} &{\bf  4}  &  5  \\
 \hline
   n(gas)\tablefootmark{a}  & 3.65(13) &8.20(12) & 1.43(12)& {\bf 4.22(11)} &{\bf 1.66(11)} & {\bf 4.47(10)}& 1.74(10)\\ 
   x(\ace)& 1.65(-4)& 1.53(-4) & 1.51(-4)& {\bf 1.20(-4)} &{\bf 1.16(-4)} & {\bf 1.18(-4)}& 1.20(-4)\\ 
   x(\ben)& 1.05(-13) &4.38(-11)& 9.70(-8)& {\bf 9.10(-7)}&{\bf 1.31(-6) }&{\bf 2.45(-7)}& 6.25(-8)\\
   N$_{tot}$(\ace) & 5.04(51) &3.43(51)& 1.74(51)& {\bf 8.52(50)}&{\bf 5.76(50)} &{\bf 3.82(50)}& 2.98(50)\\ 
   N$_{tot}$(\ben) & 3.21(42)&9.81(44) &1.12(48)& {\bf 6.46(48)}& {\bf 6.50(48)}&{\bf 7.94(47)}&1.55(47)\\ 
   Y(\ben) & 0& 0& 0.2 \% & {\bf 2.7 \% }& {\bf 3.38 \%} &{\bf 0.6 \%} & 0.1 \% \\
 %  m(\ace)& 9.8&876 & 1.79(10)&5096 &8.94(10)&1550}&1.71(11)\\
   m(\coro)\tablefootmark{a}&  1.60(21)&4.89(23) & 5.56(26)&{\bf 3.22(27)}&{\bf 3.24(27)}&{\bf 3.95(26)}&7.73(25)\\
   N$_{growth}$(\ace)& 2.89(43)&8.83(45) & 1.01(49)&{\bf 5.81(49) }&{\bf 5.85(49)}&{\bf 7.14(48)}&1.40(48)\\
 N$_{growth}$(\ace)/N$_{tot}$(\ace) &0 \%&0.0003 \%&0.58 \%&{\bf 6.83 \%}&{\bf 10.16 \%}&{\bf 1.87 \%}&0.47 \% \\
    m(AC dust)\tablefootmark{a} & 1.60(21)& 4.89(23)&{\bf 5.56(26)} &{\bf {\bf 3.22(27)}}&{\bf {\bf 3.24(27)}}&{\bf 3.95(26)}&7.73(25)\\
   m(dust)/m(gas)\tablefootmark{b}& & & $1.30\times 10^{-5}$ & $ {\bf 1.22\times 10^{-4}}$& ${\bf 1.75\times 10^{-4}}$& ${\bf 3.28\times 10^{-5}}$&$8.37\times 10^{-6}$\\
\hline
\end{tabular} 
\tablefoot{
 \tablefoottext{a}{Gas number densities are in cm$^{-3}$ and masses  in g.}
 \tablefoottext{b}{The estimated total dust-to-gas mass ratio ranges from $1.2 \times 10^{-3}$ to $5.8 \times 10^{-3}$ (see text for more detail).}
 }
\end{table*}
%--------
%-------------------------------------------------------------
%
%__________________________________

Central to this study is the formation of the first aromatic ring as a bottleneck process to AC dust synthesis in carbon stars. As seen in Figure \ref{fig6a}. the reported abundance of \ben~gradually increases with stellar radius up to $\sim$ 3 \rstar, where it reaches a maximum value to gradually decrease up to 5 \rstar. This result points to a specific radius range where AC dust has a chance to efficiently grow provided that the growing agent, in this case acetylene, \ace, is still abundant in the gas phase. To test our results, we assess the mass of AC dust that can grow from the \ben~and \ace~masses in the radius range of interest, compare the derived AC mass to the gas mass, and derive dust-to-gas mass ratios. It has been found that IRC+10216 is heavily dust-rich and has a dust-to-gas mass ratio comprised between $1.4 \times 10^{-3}$ and $4\times 10^{-3}$ based on various studies (e.g., Ivezi{\'c} \& Elitzur 1996, Groenewegen 1998, Decin et al. 2010a). In Table \ref{tab5}, we list the abundances and total masses of \ace~and \ben~ formed as a function of radius. Spherical symmetry for the inner wind is assumed to calculate the shell volume and the total number of \ace~and \ben~molecules. We assume that the growth of \ben~to coronene, \coro, proceeds through the H-abstraction-\ace-addition (HACA) mechanism (Frenklach et al. 1984) with the addition of nine \ace~molecules, and that once coronene forms, the growth of grains continues via \coro~dimerisation, coalescence, and coagulation (Cherchneff 2011b), We discard other growth mechanisms such as the polimerisation of polyynes on aromatic radical sites proposed by Krestinin et al. (2000) because the abundances of C$_4$H$_2$ are far too low to foster a significant growth of aromatic structures through this channel. Further growth via \ace~addition on the surface of grains may occur but is not considered in our simple estimate. Table \ref{tab5}~indicates that there exists a region of high dust-formation efficiency between 2.5 \rstar~and 4 \rstar. This region is characterised by high gas densities and low enough temperatures (see Table \ref{tab2}) to initiate aromatic growth, as illustrated in Figure \ref{fig6a}. Furthermore, the concentration of \ace~is sufficiently high to secure growth up to \coro~without depleting the \ace~content too much, as illustrated in Table \ref{tab5}, where $\sim 10$ \% of the \ace~species are consumed in the growth to \coro~at 3 \rstar. 

To estimate the total amount of AC dust mass formed in the inner wind and ejected at 5 \rstar, we consider a simple formalism whereby a parcel of gas moves gradually from 2 \rstar~to 5 \rstar~over a certain time span. Assuming a microturbulent velocity of between 1 and 5 km s$^{-1}$, a range characteristic of the inner wind before gas drag and acceleration by dust (Keady et al. 1988), and the stellar pulsation period of Table  \ref{tab1}, 43 pulsations (and shocks) are necessary for that parcel to reach 5 \rstar~ assuming a microturbulent velocity of 1 km s$^{-1}$, when the pulsation number drops to 9 for a microturbulent velocity of 5 km s$^{-1}$. For the modest shock velocities and moderate post-shock conditions found in the inner wind of IRC+10216, we assume that the AC dust is not destroyed in the hot post-shock gas at each shock passage, while PAH species are, but reform in the adiabatic expansion phase. Therefore, to estimate the total AC dust mass that possibly forms in the inner wind and is ejected at 5 \rstar, we assume that the total \coro~mass is converted into AC grain mass and we sum up the derived masses over the number of pulsations required to reach 5 \rstar, interpolating mass values from the data given in Table \ref{tab5}. We obtain a total dust-to-gas mass ratio that spans the range $1.2 \times 10^{-3}-5.8 \times 10^{-3}$. When compared to values derived from observations ($1\times 10^{-3}$ to $4 \times 10^{-3}$), these numbers are satisfactory and point to a specific region in the inner wind where carbon dust grows from PAHs and graphene sheets. Our simple derivation is based on a 100 \% growth efficiency of \ben~to \coro~and a 100 \% conversion efficiency of \coro~in AC dust grains. These assumptions clearly maximise the AC dust mass value at 5 \rstar. However, owing to the large \ace~reservoir available in this region, an additional growth process that are not considered in the present derivation includes the addition of \ace~molecules at the surface of graphene sheets, a mechanism that would add mass to the final carbon dust budget of the inner wind. Without pointing to a specific value for the dust-to-gas mass ratio, our derived ratios closely agree with the dust-to-gas mass ratios derived from observations and are indicative that AC dust formation at these specific radii efficiently proceeds in the inner wind of IRC+10216. 

%--------------
\subsection{Other dust precursors: Carbides and sulphides}
\label{other} 
\subsubsection{SiC$_2$ and SiC}
\label{sic}
%-------------------------------------------------------------
   \begin{figure}
   \centering
   \includegraphics[angle=0,width=8.9cm]{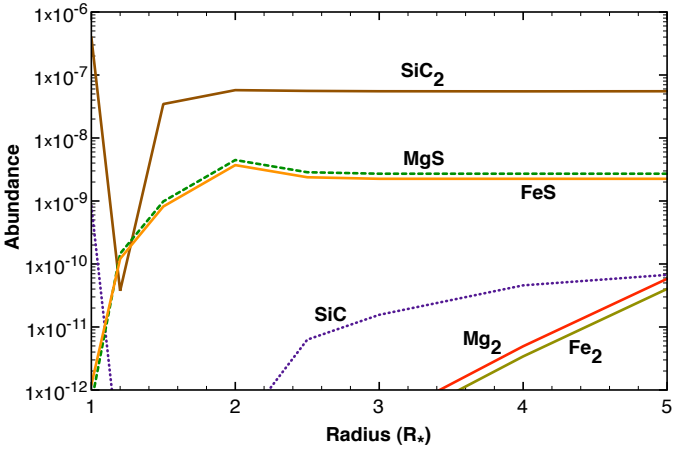}
      \caption{Abundances (with respect to H$_2$) of SiC and SiC$_2$, and Mg- or Fe-bearing species involved in the formation of metallic clusters as a function of radius. Abundances derived by thermodynamic equilibrium at r$_s$ are moved to radius 1 \rstar~for clarity. Except for SiC$_2$ and metal sulphide cluster precursors, abundances are quite low in the dust formation zone, and pure metal clusters do not form from the gas phase.
              }
         \label{fig7}
   \end{figure}
%
%__________________________________
Silicon carbide (SiC) dust has long been observed in the wind of carbon stars through its transition at 11  \micron~(Treffers \& Cohen 1974, Speck et al. 1997) and studies on meteorites have confirmed a AGB origin for some of the pre-solar SiC inclusions (Zinner 2007). In the laboratory, the synthesis of SiC nanoparticles is produced using various experimental methods (e.g., laser-induced pyrolysis of gas-phase mixtures of silane, SiH$_4$, and hydrocarbons)  and SiC grains are observed to form in the temperature range $1600-2000$ K. These temperatures are encountered in the inner wind, and support the hypothesis that SiC forms from the gas phase by chemical kinetic processes similar to those active in SiC synthesis in the laboratory. Despite the nucleation processes not being fully understood and characterised, a few gas phase species have been identified as intermediates in the nucleation of SiC particles and include Si, C$_2$, and cyclic SiC$_2$ (Fantoni et al. 1991). Silicon dicarbide, SiC$_2$, has been detected in the inner wind of IRC+10216 by mm interferometry with abundances that range from $ < 5 \times 10^{-8}$ (Gensheimer et al. 1995) to $5 \times 10^{-7}$ (Lucas et al. 1995). Recent observations with HIFI onboard Herschel indicate a SiC$_2$ abundance with respect to H$_2$ of $2 \times 10^{-7}$ in the inner wind (Cernicharo et al. 2010a). We thus assume that the presence of SiC$_2$ in the dust formation zone reflects the nucleation and condensation of SiC grains at high temperatures and densities and the role of intermediate plays by this species. 

There exist no documented reaction rates for SiC$_2$ formation and we rely on the isovalence of silicon with carbon to derive reasonable rates for specific processes. We also consider the identified nucleation routes for SiC clusters and the formation of SiC$_2$ and (SiC)$_2$ clusters according to Erhart \& Albe (2005). The main production process for SiC$_2$ in the gas phase is
%-------
\begin{equation}
\label{sic1}
   {\rm SiC + SiC} \rightarrow {\rm SiC_2 + Si,}
\end{equation}
%------
where destruction is commanded  by the reverse reaction in Eq. \ref{sic1}~and thermal fragmentation. We also assume that two SiC molecules react to form the SiC dimer (SiC)$_2$. The results for SiC and SiC$_2$ abundances are shown in Figure \ref{fig7}. The shock at r$_s$ destroys both the SiC and SiC$_2$ initially present in the photosphere under TE, but SiC reforms in the post-shock gas at 1.5 \rstar~and form SiC$_2$ via the reaction in Eq. \ref{sic1} and SiC dimers. At 5 \rstar, the SiC$_2$ abundance value agrees well with that derived by Cernicharo et al. (2010a), indicating that the molecule may be regarded as a by-product of the condensation of SiC clusters at small radii. The model shows that both SiC$_2$ and (SiC)$_2$ form in large amounts as early as 1.5 \rstar, and well before the aromatic formation zone (2.5 \rstar~- 4 \rstar) discussed in \S \ref{arom}. SiC clusters thus represent a high temperature condensate population independent of the synthesis of AC dust. We discuss in more detail the consequences of this situation for wind acceleration in \S \ref{dis}.  

\subsubsection{MgS and FeS}
\label{mgs}

A strong 30 \micron\ emission band has been observed in carbon-rich evolved stars at various stages of their evolution, including AGB and post-AGB stars, and planetary nebulae. Observation of the band was also reported in IRC+10216 and ascribed to solid magnesium sulphide, MgS, for which a low radiative temperature comprised between 100K and 450 K was derived (Goebel \& Moseley 1985, Szczerba et al. 1999, Hony et al. 2002a). A band at 23 \micron~was observed in two carbon-rich planetary nebulae and FeS in the form of troilite was proposed as a possible carrier (Hony et al. 2002b). FeS is also responsible for the 23 \micron~band detected in proto-planetary discs (Keller et al. 2002). However, there is no observational evidence that this band is present in the spectral energy distribution of C-rich evolved stars in general, including IRC+10216. Chemical models assuming TE in the inner wind predict that both MgS and FeS condense in carbon-rich environments (Lattimer et al. 1978, Lodders \& Fegley 1999). Keeping in mind that dust formation is not an equilibrium process in stellar outflows, we test the formation of MgS and FeS molecules as the initial gas-phase precursors of MgS and FeS grains in the inner wind. We also consider the formation of Mg- and Fe- bearing molecular species and clusters. They include the pure iron and magnesium small clusters, Fe$_2$ and Mg$_2$ respectively, although there exists no observational evidence of pure metal clusters in AGB environments. However, pure iron grains are often proposed as a dust component of O-rich AGB winds to account for the required near-IR opacity necessary to accelerate the wind (Woitke 2006). We also consider the hydrides MgH and FeH, and the gas-phase precursors to metal oxides MgO and FeO. 

Atomic magnesium is an alkaline earth metal that primarily reacts with oxygen compounds (e.g., H$_2$O, O$_2$, NO, O$_3$) to form magnesium oxides. Although reactions with sulphur-bearing compounds are not documented, we expect similar types of reaction between Mg and SO to those between Mg and O$_2$, owing to the isovalence of sulphur with oxygen.  We therefore assume that Mg will react with SO according to the reaction

\begin{equation}
\label{mg1}
   {\rm Mg + SO} \rightarrow {\rm MgS + O}
\end{equation}
%--------
Using again the isovalence of sulphur with oxygen, we assume that a reaction similar to the reaction in Eq. \ref{mg1} triggers the formation of molecular FeS. For the formation of iron monoxide, FeO, and magnesium monoxide, MgO, reactions of atomic Fe and Mg with O$_2$ are considered. These processes have been extensively studied (e.g., Akhmadov et al. 1988) and their rates well-documented. For both sulphides, we also consider the following radiative association reaction as a possible production channel 

\begin{equation}
\label{mg2}
   {\rm X + S} \rightarrow {\rm XS + h\nu,}
\end{equation}
%----
where $X = $ Mg or Fe. Studies by Kimura et al. (2005a, 2005b) that explore various formation routes to MgS and FeS in the laboratory show very efficient synthesis from the reaction of the two Mg (Fe) and S gaseous phases in gas flash evaporation methods, and support the occurrence of the reaction in Eq. \ref{mg2} and its termolecular analogue process.  
The formation of Fe$_2$ is described by the reaction

\begin{equation}
\label{fe1}
   {\rm Fe + Fe + M} \rightarrow {\rm Fe_2 + M,}
\end{equation}
%----
where $M$ is the gas collider. A rate for the reaction in Eq. \ref{fe1} was derived by Giesen et al. (2003) in their study of pure iron cluster formation at high temperatures. The radiative association reaction between two Fe atoms is also included. Similar types of processes and rates are considered for the synthesis of Mg$_2$. We assume that the reverse reactions of all chemical pathways are the only destruction processes operating on Fe$_2$ and Mg$_2$. 

Abundances with respect to H$_2$ for these species are shown in Figure \ref{fig7}, except for FeO and MgO which have negligible abundances in the inner wind ($x$(MgO, FeO) $\sim 1\times 10^{-18}$). Apart from MgS and FeS, Mg- and Fe-bearing species have very low abundances. Specifically, Fe and Mg do not form pure metal clusters as Fe$_2$ and Mg$_2$ have low abundances in the dust formation zone. According to our model, most of the Mg and Fe initially present in the photosphere and at r$_s$ stays in atomic form. A moderate amount of Mg and Fe is first included in the hydrides MgH and FeH (see Figure \ref{fig2}) but are preferentially included into  MgS and FeS at $r \ge$ 2 \rstar~from the reaction of Mg and Fe with SO following the reaction n Eq. \ref{mg1}. Similar to H$_2$O, sulphur monoxide formation is induced by the release of oxygen atoms in the collisional dissociation of CO in the hot post-shock gas. 

An upper limit to the total mass of solid MgS and FeS produced is derived assuming that all gas-phase MgS and FeS are depleted into clusters and grains once the gas has reached the temperature regime (T $< 450$ K) derived by Goebel \& Moseley (1985). The mass limit for both MgS and FeS is at most $\sim 0.2 \%$ of the carbon dust mass formed between r$_s$ and 5 \rstar. When modelling the spectral energy distribution of IRC+10216, Ivezi{\'c} \& Elitzur (1996) derive a dust composition where MgS accounts for less than 10 \% by mass. For the carbon-rich post-AGB star HD 56126, Hony et al. (2003) found that a MgS mass of $\sim$ 2 \% of the carbon dust mass is necessary to account for the 30 \micron~band flux. One would thus expect the MgS mass to represent at most a few percent of the AC dust mass in the inner wind of carbon stars. This required mass is higher than our upper limit by a factor of ten. This discrepancy may be the result of several uncertainties in the sulphur chemistry or may point to either a nucleation process for MgS that does not occur in the gas phase or to a different carrier for the 30 \micron~band. These various aspects are discussed in \S \ref{dis}. 
%This discrepancy may be indicative of different effects. Firstly, atomic sulphur is essentially depleted in SiS at r$_s$ and our modelled SiS abundances are larger than those observed by a factor of $\sim 3$. The chemistry of SiS being poorly characterised, it is possible that the model overestimate SiS abundances, resulting in less atomic sulphur available for forming SO and therefore MgS through Reaction \ref{mg1}. Secondly, if MgS grains are responsible for the 30 \micron~band, they cannot form directly from gas phase precursors in the dust formation zone. Other formation mechanisms for MgS at relatively low temperatures must operate, including formation on the surface of pre-existing dust grains. Although this scenario has been often proposed, it is not clear from the present study how MgS clusters can grow from the gas-surface chemistry because atomic sulphur, the growing agent with Mg, is chiefly depleted in SiS and CS at short stellar radii and has a negligible abundance at 5 \rstar. 

Gas-phase FeS follows an abundance trend similar to MgS, as seen in Figure \ref{fig7}, because we assumed that the FeS and MgS chemistries were alike. Both Fe and Mg are mainly in atomic form and have very similar abundances at TE in the photosphere. Both solid MgS and FeS possess identical clustering structures going from (XS)$_2$ (X = Mg, Fe) with its planar rhombic structure to (XS)$_4$, which quickly reaches a distorted cubic structure. However, MgS clusters are unstable in O-rich environments, contrarily to FeS clusters. It was proposed by Begemann et al. (1994) that composite solid sulphides including both Mg and Fe could satisfactorily reproduce the 30 \micron~band in IRC+10216. In particular, a magnesium-iron sulphides whose composition ranged from Mg$_{0.9}$Fe$_{0.1}$S to Mg$_{0.5}$Fe$_{0.5}$S provided the closest matches to the band. Such a carrier could indeed be synthesized in the dust formation zone of IRC+10216 in view of the Fe- and Mg-bearing species that form, and the presence of both gas-phase MgS and FeS. 

As pointed out before, most of the Mg and Fe in the inner wind region is in atomic form. High resolution observations of optical absorption lines of several metals by Mauron \& Huggins (2010) for IRC+10216 also appear to detect atomic metals in the gas phase, in direct contradiction with TE condensation models for which all metals are depleted in a solid phase once the condensation temperature of the solid is reached in the wind. However, the derived abundances for iron and calcium atoms highlighted some degree of depletion relative to the solar abundance values. According to this study, the depletion cannot result from the trapping in a molecular phase, as corroborated by the present results where the abundances of metal-bearing species are always less than $1\times 10^{-6}$. The partial depletion of iron and calcium must thus result from either the incorporation of free-flying Fe and Ca atoms in the inner wind during the AC dust condensation process at $r \leq 5$, or the adsorption of these atoms on the surface of dust grains at the lower gas temperatures encountered at larger envelope radii. 
%------
\subsection{Line variability with time}
\label{var}
%-------------------------------------------------------------
   \begin{figure}
   \centering
   \includegraphics[angle=0,width=8.9cm]{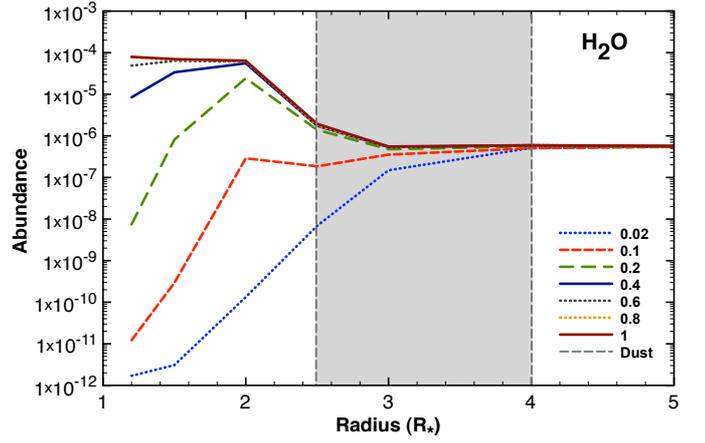}
      \caption{H$_2$O abundances (with respect to total gas) as a function of radius and pulsation phase (legend). Phases 0.8 and 1 are indistinguishable. The grey shaded area corresponds to the zone where benzene forms and aromatics grow to form carbon dust (see \S~\ref{arom}).
              }
         \label{fig8}
   \end{figure}
%
%__________________________________
Our model predicts that specific molecular abundances have a strong time-dependence. Some molecular species indeed show a strong variability in their abundances as a function of time or phase of the pulsation period. The abundance of water is a good example of such a variability and its abundance with respect to H$_2$ as a function of radius and pulsation phase is shown in Figure \ref{fig8}. At 2 \rstar, H$_2$O abundances span almost six orders of magnitude over one pulsation period (P = 650 days), and this variation should be reflected in the intensity of its high-energy transitions. In these deep layers, the transitions are mainly pumped by IR radiation. Apart from the intrinsic variability of the stellar flux with pulsation phase that affects all molecules, the large variations in abundances should have some impact on the high excitation-line fluxes. These changes in abundances are the consequence of the post-shock non-equilibrium chemistry and the destruction of molecular species in the hot post-shock gas at early phases. 

Destruction is more or less severe depending on the species and not all molecules behave like H$_2$O. For example, with its strong molecular bond, CO does not experience such variations and shows a rather constant abundance distribution with radius and pulsation phase. Some species (e.g., SiO), show a time variation in abundance that increases in amplitude from large radii to small radii, but where the variation amplitudes do not span more than one order of magnitude. Finally, other molecules (e.g., SiC$_2$) show average amplitude variations (of three orders of magnitude) at small radii but reach time-independent abundances deep inside the inner wind ($\sim$ 2.4 \rstar). Although the impact of time-varying abundances  on the line fluxes is difficult to quantify without a proper radiative transfer model, we would expect some species to show little line variability (e.g. CO) but others to be prone to moderate variability (e.g., SiO) and undergo large line-flux changes with time (e.g., H$_2$O). Clearly water is an excellent tracer of time variability of high-energy molecular transitions and as such, an excellent indicator of shock activity and shock-induced chemistry in the dust  formation zone. 

\section{Summary and discussion}
\label{dis}

We have modelled the inner wind of the carbon star IRC+10216  assuming the periodic propagation of pulsation-driven shocks between 1 \rstar~and 5 \rstar~and considering a complete gas-phase chemistry that encompasses several chemical families. These shocks trigger a non-equilibrium chemistry in the hot post-shock gas that leads to the formation of molecules and dust precursors. The study points to the following new results and trends applicable in general to carbon stars: 
\begin{itemize}
\item The model confirms the presence of a group of molecules, namely CO, HCN, SiO, CS, and H$_2$O, that efficiently form between 1 \rstar~and 5 \rstar. These species are expected to be present, albeit in different quantities, in the inner wind of all AGB stars, regardless of their C/O ratio, as already proposed by Cherchneff (2006, 2011a). The derived abundance values agree well with available observations. In particular, the dissociation of CO by collisions in the immediate post-shock gas triggers the formation of atomic O, OH, SiO, and H$_2$O.  
\item We have found that some hydrides form a new category of abundant and stable species in the inner wind. These include AlH,  HCl, and HF.   
Other hydrides do form in large amounts at r$_s$ but are rapidly converted in the dust formation zone into chlorine-bearing species, leaving the wind acceleration region with low abundances. The most abundant hydrides will be released to the intermediate envelope and participate in an active chemistry at larger radii.
\item Once formed from the reaction of H$_2$ with Cl, HCl is the production agent of other Cl-bearing species such as AlCl, NaCl, or KCl. The formation chemistry of chlorine-bearing species is thus independent of the C/O ratio of the stellar photosphere, and Cl-bearing molecules including HCl, NaCl, and KCl should also be present in O-rich AGB and supergiant stars, albeit with different abundances. AlCl is expected to have a lower abundance in the dust formation zone of O-rich sources because of the Al depletion in gas- or solid-phase metal oxide species. 
\item There exists a specific zone that extends from 2.5 \rstar~to 4~\rstar, where the closure of \ben~ring occurs through the recombination of two propargyl radicals, with a \ben~abundance peaking at 3 \rstar. The available \ace~abundances in the gas phase are high enough to secure growth to PAHs such as \coro. These large PAHs will consecutively coalesce and coagulate to form AC dust. The estimated total dust-to-gas mass ratio spans the value range $1.2 -5.8 \times 10^{-3}$ and closely agrees with existing values derived from observations of IRC+10216. Within this zone, SiC$_2$ molecules efficiently form as by-products of the synthesis of SiC clusters. Some MgS and FeS species are also produced in the gas phase but their abundances are too low to account for the 26-30 \micron~emission band. 
\item The shock-induced scenario predicts a time-variability for some molecular abundances over a pulsation period (e.g. H$_2$O and SiO) that should induce a time variability in their high excitation line fluxes. Other species show either negligible changes in abundances (e.g., CO), or small changes that do not affect the molecular line intensity (e.g., SiC$_2$). This predicted time variability is the direct result of the destruction of species in the hot shocked gas layers. Observations of the high-energy transitions of these specific species at different epochs of the pulsation period would help us to confirm the predicted time variability, and assess the impact of shocks on the gas chemistry. 
\end{itemize}

% dust PAHs, SiC, and MgS

As reported in \S~\ref{arom}, we have found that the formation of PAH molecules, and both their coalescence and growth to AC dust  take place in a specific radius range. The growth of benzene, \ben, to coronene, \coro, via the HACA mechanism is expected to consume a large part of the benzene rings synthesised at these radii. At radii larger than 4 \rstar, inspection of Table \ref{tab4} shows that some benzene rings still form and can grow to larger aromatic species as the \ace~reservoir is still large. However, the lower gas densities and temperatures should hinder the coalescence of PAHs to form large graphene structures. A population of free-flying PAHs not incorporated in AC dust should thus be expelled to larger radii once the wind is fully acccelerated. The so-called unidentified infrared bands are observed in carbon stars that are part of binary systems (Speck \& Barlow 1997, Boersma et al. 2006). For the carbon star TU Tau, the UV radiation field of the blue companion could excite the aromatics present in the carbon star wind. These excited PAHs might include the free aromatics that are synthesised beyond the aromatic formation zone highlighted in this study. 

We have also found that SiC dimers form at $\sim 1.5$ \rstar, i.e., far smaller radii than the aromatic growth region, which implies that there is an independent population of SiC clusters at small radii. Owing to the  extinction properties of SiC dust and a significant decrease in the Planck mean of its extinction efficiency for temperatures corresponding to the effective temperatures of AGB stars, the dust experiences an inverse greenhouse effect for a radiation field characteristic of carbon stars (Gilman 1974, McCabe 1982, Yasuda \& Kozasa 2011). Since the pressure force acting on dust grains is directly proportional to the Planck mean of the extinction efficiency, most of the acceleration of the wind is provided by AC dust grains (Cherchneff et al. 1991). On the other hand, the inner SiC cluster population should experience a minor radiation pressure force and lags behind the AC clusters when expelled to larger radii. This situation may be reflected in the results of studies of meteorites. Pre-solar SiC grains bearing the isotopic fingerprint of the AGB s-process are not included in graphite spherules that have an AGB origin but form a separate pre-solar grain population (Hynes et al. 2007). This isolation of the SiC presolar grains may directly result from the non-equilibrium chemistry in the post-shock gas that produces two main dust populations, namely SiC and AC grains, at two distinct positions in the wind acceleration zone. 

No firm conclusions about the production of MgS or Mg-FeS dust grains in IRC+10216 can be drawn from the present results, as they can be interpreted in many ways. Firstly, the chemical model may underestimate the MgS and FeS abundances by a factor of ten or more because too much atomic sulphur is trapped in SiS, as indicated by the slightly higher SiS abundances listed in Table \ref{tab3} relative to those derived from observations. A small amount of S not locked in SiS would result in higher SO abundances and in a larger amount of MgS and FeS owing to the large reservoirs of free atomic Mg and Fe in the inner wind. The MgS and FeS clusters would then be produced from the gas phase at r $\ge$ 2 \rstar, and the resulting estimated MgS dust mass could reach the few percent of AC dust mass necessary to explain the 30 \micron~band. That the present model forms gas-phase FeS and MgS with similar efficiencies indicate that it may form composite Mg-Fe sulphide clusters instead of pure MgS clusters, as proposed by Begemann et al. (1994). According to Kimura et al. (2005a, 2005b), a MgS (FeS) nucleation from the gas phase produces spherical cubic MgS (FeS) clusters instead of the elongated network-like grains that form when gas-surface reactions are involved in the nucleation process. Hony et al (2002b) studied the effect of dust shape and temperature on the band shift in wavelength with a shift towards 26 \micron~when spherical and hot grains were considered. The emission band in IRC+10216 clearly peaks around 27 \micron~in the ISO spectrum, pointing to a possible formation pathway from gas-phase chemistry, as we have described in the present study. Secondly, if the MgS (FeS) abundances are indeed low in the dust formation zone, they point to: 1) a synthesis mechanism for MgS or Mg-FeS involving gas-surface processes, or 2) alternative carriers for the 30 \micron~band. However, if MgS grain formation occurs at lower temperatures and on the surface of already produced dust grains (see the comprehensive studies by Men'shchikov et al. (2001) and Zukhovska \& Gail 2008), their growth is hindered by the lack of available atomic sulphur, which is chiefly depleted in SiS and CS in the dust production zone. 
Therefore, a formation scenario involving surface chemistry would also require a mechanism that could return sulphur to the gas phase just after the acceleration of the outflow. Another explanation is that MgS is not the carrier of the 30 \micron~band. A critical assessment of all previous MgS studies was made by Zhang et al. (2009) who pointed out that the mass of MgS derived from the emission at 30 \micron~violated the available abundances of Mg and S in the stellar atmospheres, owing to the use of improper optical constants for MgS in the optical and UV wavelength domains. Alternative solids have been proposed (e.g., hydrogenated amorphous carbon, HAC, Grishko et al. 2001). A fresh reinvestigation of the carrier of the band in IRC+10216 would be extremely useful coupled to observations of high energy transitions of sulphur-bearing species to constrain the sulphur reservoir in the dust formation zone.

%As stated in \S~\ref{prev}, two scenarios now prevail for explaining the formation of H$_2$O at such small radii: the photo-dissociation of CO isotopes by the interstellar UV radiation field penetrating deep in the wind (Decin et al. 2010, Ag{\' u}ndez et al. 2011) and the shock-induced non-equilibrium chemistry presented in Cherchneff (2001) and discussed in the present study. As explained by Neufeld et al. (2011a), the photo-dissociation scenario assumes that $^{12}$C$^{16}$O shields itself in the deep layers and that oxygen is released from the UV dissociation of CO isotopes, namely $^{13}$CO, C$^{17}$O and C$^{18}$O. The release of free oxygen isotopes to form H$_2$O should thus result in an enhancement of  the H$_2$$^{17}$O/ and H$_2$$^{18}$O abundances compared to H$_2$$^{16}$O, leading to an enhancement of the H$_2$$^{17}$O/H$_2$$^{16}$O and the H$_2$$^{18}$O/H$_2$$^{16}$O ratios compared to the isotopic ratios of  $^{17}$O/$^{16}$O and $^{18}$O/$^{16}$O. By contrast, the destruction of CO by collisions in the post-shock gas (Cherchneff 2011) does not impact the abundances of water isotopologues. Thus, the observation of high energy transitions of these isotopologues are crucial to discriminate between the two scenarii. 

% various amounts of water in carbon stars 
Finally, the results presented in this study are not unique to IRC+10216 and similar trends should apply to other carbon stars as well. For the specific case of water, H$_2$O has now been detected in several carbon stars (Neufeld et al. 2011b) where formation processes similar to those described in Cherchneff (2011a) and in the present study take place. However, the H$_2$O abundance certainly varies from source to source depending on the various parameters that are entangled with its formation in a complex way. For example, the shock strength affecs both the destruction of molecules and the creation of free atomic oxygen. Therefore, one would expect lower water abundances to be created by a mild shock than a strong shock, but the opposite actually occurs. We have modelled the chemistry induced by a 10 km s$^{-1}$ shock at r$_s$ and compared our results with those of the 20 km s$^{-1}$ shock, finding that more SiO and H$_2$O molecules were produced. Because less CO is destroyed by a mild shock, the SiO formation depends on the reaction of Si with CO. Therefore, the OH radical is free to form H$_2$O, and combined with the less efficient destruction of molecules in the post-shock gas, more water is formed. Hence, the many parameters affecting the formation of water in carbon stars include the shock strength as well as the chemical composition, both the photospheric gas density and temperature, the gas-phase chemistry of the Si and S chemical families, and the amount and type of dust that forms. Water abundances is thus expected to vary greatly in carbon stars despite its synthesis in these objects having been proven both observationally and theoretically. A similar conclusion may be drawn for other species such as SiO. Combined observations of several high excitation transitions of H$_2$O and SiO molecules would be very instructive in this regard to more clearly understand the chemical processes responsible for the formation of O-bearing species and characterise the water content of carbon stars on a global scale.   

%It should be mentioned that the photodissociation scenario should result in the enhancement of other O-bearing molecular isotopologues, e.g., isotopologues of SiO. As seen in section \S~\ref{prev}, both SiO and H$_2$O are formed from similar processes that involve the reaction of OH with Si and H$_2$, respectively. Therefore, Si$^{17}$O and Si$^{18}$O abundances should be enhanced with respect to Si$^{16}$O if the isotope-selective photodissociation prevails in the inner wind. Observations of H$_2$O and SiO isotopologues are thus of prime importance to discriminate between the two proposed mechanisms. 

\begin{acknowledgements}
 The author thanks the two anonymous referees for their useful comments that helped to improve the manuscript, A. Tielens for constructive remarks, and D. Gobrecht for providing the TE calculation estimates.
\end{acknowledgements}

\begin{appendix}
\begin{table*}
\caption{The chemical reaction network ordered by type of reactions. The reaction rate is given in the Arrhenius form A x (T/300)$^n$ x exp(- E$_a$/T), where T is the temperature in Kelvin. The related parameters are given in each column as follows: Column 1 -  reaction number; Column 2  - reactants and products; Column 3 - A parameter (in cm$^3$ s$^{-1}$, cm$^6$ s$^{-1}$, and s$^{-1}$ for bimolecular, trimolecular, and unimolecular processes, respectively); Column 4 - n parameter;  Column 5 - activation barrier E$_a$ in Kelvin: Column 6 - rate reference.}          
% title of Table
\label{tab1ap}     
\centering                          % used for centering table
\begin{tabular}{llclcccl}       
\hline\hline                
 &&&TERMOLECULAR &&&\\
3B1 &  H   +  H  + H$_2$  & $\longrightarrow$& H$_2$  + H$_2$ & $ 8.85\times 10^{-33}$& -0.6  &0 & NIST \\
3B2  & H     + H    + H   & $\longrightarrow$ &   H$_2$   + H    &   $8.82\times10^{-33}$ &     0   &     0    &NIST\\
3B3 &  H     + H    + He   & $\longrightarrow$&   H$_2$   + He   &  $4.96\times10^{-33}$ &    0 &       0   &  NIST\\  
3B4  & H     + O    + M   & $\longrightarrow$&   OH   + M   &   $ 4.36\times10^{-32}$&    -1.0    &    0  &  NIST\\
\end{tabular}
\end{table*}
\end{appendix}
\end{document}